\documentclass[11pt]{article}
\usepackage{rotating}

\usepackage{graphicx}
\begin{document}
\title{\bf Peculiarities of the Abundances of Neutron-Capture Elements
in Galactic Open Clusters}

\author{{V.\,A.~Marsakov, M.\,L. Gozha, V.\,V.~Koval', and L.\,V. Shpigel'}\\
{Southern Federal University, Rostov-on-Don, Russia}\\
{e-mail:  marsakov@sfedu.ru, gozha$_{-}$marina@mail.ru, vvkoval@sfedu.ru}}
\date{accepted \ 2016, Astronomy Reports, Vol. 60, No. 1, pp. 61-72}

\maketitle

\begin {abstract}

The properties of the relative abundances of rapid and slow 
neutron-capture elements are studied using a catalog containing 
spectroscopic abundance determinations for 14~elements produced in
various nuclear-synthesis processes for 90~open clusters. The 
catalog also contains the positions, ages, velocities, and 
elements of the Galactic orbits of the clusters. The relative 
abundances of both $r$-elements (Eu) and $s$-elements (Y, Ba, La, and Ce)
in clusters with high, elongated orbits and in field stars of
the Galactic thin disk display different dependences on metallicity, 
age, Galactocentric distance, and the elements of the Galactic orbits, 
supporting the view that these objects have different natures. In young
clusters, not only barium, but also the three other studied $s$-elements
display significantly higher relative abundances than field stars of 
the same metallicity. The relative abundances of Eu are lower in 
high-metallicity clusters (${\rm [Fe/H]} > -0.1$) with high, elongated 
orbits than in field giants, on average, while the [Eu/Fe] ratios 
in lower-metallicity clusters are the same as those in field stars, 
on average, although with a large scatter. The metallicity dependence 
of the [O, Mg/Eu] ratios in clusters with high, elongated orbits
and in field stars are substantially different. These and other 
described properties of the Eu abundances, together with the 
properties of the abundances of primary $\alpha$-elements, can be understood
in a natural way if clusters with high, elongated orbits with 
different metallicities formed as a result of interactions
of two types of high-velocity clouds with the interstellar medium 
of the Galactic disk: low-metallicity high-velocity clouds that formed 
from ``primordial'' gas, and high-metallicity clouds with intermediate 
velocities that formed in ``Galactic fountains''.

\end{abstract}

{\em open star clusters, chemical composition, kinematics, Galaxy 
(Milky Way)}.

\maketitle

\section{Introduction}

This paper continues our systematic study of the
chemical compositions of open clusters, motivated by
the existence of clusters with anomalously elongated
orbits extending high above the Galactic plane or with
low metallicities, which are most likely characteristic
of old Galactic subsystems. One proposed explanation
for these properties is that the population of open
clusters is not uniform, and includes clusters formed
partially from interstellar material that fell in from
outer volumes of the Galaxy [1-4] (see also references
therein). The chemical compositions of such clusters
should deviate from those for field stars of the thin
disk. To provide a basis for studies of the chemical and
spatial-kinematic properties of samples of clusters,
we have compiled a catalog of 346 clusters based on
data published over the past 23 years, including the
positions, metallicities, ages, and relative abundances
of 14 elements for 90 open clusters, for 82 of which
the elements of their Galactic orbits have been calculated
[5, Table 2]
\footnote{A full version of this table is available in electronic 
form at the site ftp://cdsarc.u-strasbg.fr/pub/cats/J/AZh.}

Our analysis of the relative abundances of $\alpha$\-elements
in clusters with various kinematic characteristics
in [5] supported the existence of different types of
clusters with different natures. In particular, we found
that the relative abundances of primary $\alpha$-elements
([O, Mg/Fe]) in clusters with high, elongated orbits
and field stars of the thin disk display different dependences
on metallicity, age, Galactocentric distance,
and distance from the Galactic plane. We believe that
the observed differences confirm that open clusters
with high, elongated orbits and field stars of the thin
disk have different natures.

The conclusion of [2] that some clusters formed as
a result of interactions of high-velocity, metal-poor
clouds with the interstellar material of the Galactic
thin disk is also supported by the results of~[5].
The relative abundances of $\alpha$-elements for clusters
with high, elongated orbits and metallicities ${\rm [Fe/H]} <
-0.1$ were found to be lower than the typical values for
field stars. This can be explained if the high-velocity
clouds that gave rise to them were formed of interstellar material
from regions where the star-formation
rate and/or the masses of Type II supernovae were
lower than near the Galactic plane. On the other
hand, the higher relative abundances of primary 
$\alpha$-elements observed for higher-metallicity clusters with
high, elongated orbits, compared to those for field
stars, suggests that they formed as a result of interactions
between the interstellar medium of the Galactic
disk and metal-rich clouds having intermediate
velocities, which are currently believed to form from
returning gas in so-called ``Galactic fountains''.

Some studies have shown that open clusters and
field stars also have different relative abundances of
neutron-capture elements. For example, D'Orazi
et\,al. [6] and Maiorca et\,al. [7] found that, in
contrast to field stars, the relative abundances of
slow neutron-capture elements ($s$-elements) [$s$/Fe]
in about 20 clusters increase with decreasing age.
This could come about due to an increased contribution
to enrichment of the interstellar medium by low-mass
($<1.5 M_{\odot}$) giants on the asymptotic branch
with time [8]. This result is confirmed Mishenina et
al. [9, 10], who found an appreciable excess of the
relative barium abundances in several young open
clusters, while there was no excess in the relative
abundances of the other two $s$-elements yttrium and
lanthanum.

Most studies have found clusters to have not only
barium relative abundance excesses compared to field
stars, but also an increase in [Ba/Fe] with decreasing
age, while no age dependence (see, e.\,g.,~[11]), or
even a slight growth in their relative abundances with
age (see, e.\,g., [12]), is observed for other $s$-elements.
D'Orazi et\,al. [13] also found appreciable relative
barium excesses in three young associations (which
also contain open clusters), while other $s$-elements
displayed approximately solar [$s$/Fe] values. However,
it turns out that some field stars---Cepheids ---
also display excesses of the relative abundances of $s$-elements
[14], with the highest excesses being those
for barium.

Here, we continue our comparative analysis of
the properties of the relative elemental abundances of
open clusters and field stars of the thin Galactic disk,
and also of different groups of cluster populations,
this time looking at rapid ($r$) and slow ($s$) neutron-capture
elements. Our catalog contains data on the
chemical compositions of 90 open clusters taken from
109 primary references, a list of which is provided.
The catalog includes relative abundances of both $s$-
(Y, Ba, La, Ce, Nd, and Zr) and $r$-(Eu) elements.
Yttrium abundances are included for 59, zirconium
for 52, neodymium for 29, cerium for 38, barium for
73, lanthanum for 55, and europium for 47 clusters.
The mean number of studied stars in each cluster is
seven. The abundances were determined using only
one star in 16 cases; in only five clusters was there
only one determination, and [el/Fe] values were available
in more than one study in the other cases. For
clusters with two or more abundance determinations
for a given element, we calculated the mean weighted
abundance (see [5] for more detail). Inspection of
the [el/Fe] values from different studies indicated that
the scatter of the values for barium, lanthanum, and
europium was $\sigma{\rm[el/Fe]} = 0.12 \pm0.02$, while the scatter
for yttrium and cerium was somewhat smaller [5,
Table 1]. For comparison, we consulted catalogs of
abundances for these same elements in 212 dwarfs,
171 red giants, and 221 field Cepheids of the Galactic
thin disk. We also calculated the ages and elements
of the Galactic orbits for the red giants. More detail
about the parameters of these comparison stars can
be found in [5, 14].


\section {METALLICITY DEPENDENCES OF THE RELATIVE ABUNDANCES
OF NEUTRON-CAPTURE ELEMENTS}

{\bf $s$-Elements}. According to theoretical computations,
most of these elements are produced in the
cores of giants on the asymptotic branch with masses
$M < 4M_{\odot}$, and are then ejected into the interstellar
medium in later stages of the star's evolution [15].
The peak production of these elements is at ${\rm [Fe/H]} < -0.2$, 
and [$s$/Fe] decreases with further increase in
the metallicity. In addition, some quantity of $s$-element
nuclei are produced in the cores of massive
stars ($M > 8M_{\odot}$) [16, 17]. (For more detail on the
sources of neutron-capture elements, see [14] and
references therein.) Since $s$-elements are synthesized
in stars with the same masses as Type Ia supernovae,
which eject iron-group elements when they explode,
the interstellar medium is enriched in both types of
elements simultaneously. This means that variations
in the relative abundances of $s$-elements with increasing
metallicity in a stellar-gas system could be
due exclusively to properties of the stars producing
them.

Figures~1a-1d present the metallicity dependences
of the relative abundances of four $s$-elements
(Y, Ba, La, and Ce) for open clusters and field giants
of the thin disk. Although the theoretical dependence
of the output of light $s$-elements (Y) on the metallicity
of the stars in which they are synthesized is somewhat
different than the corresponding dependence for heavy
$s$-elements (Ba, La, Ce), we included Y in this
analysis, since the relative abundances of this element
were used in [9] in a comparison with the behavior
of the barium abundance in young open clusters.
Fig.~1a shows that the relative abundances of yttrium
display appreciable scatter, for both the clusters and
field stars. In order to avoid introducing distortions
associated with the theoretical predictions of the
outputs of the various elements, we constructed linear
regression fits for all the $s$-elements for the range
${\rm [Fe/H]} > -0.2$, where, in fact, most open clusters
and field stars of the thin disk are located. The slope
of the dependence $d{\rm [Y/Fe]}/d{\rm [Fe/H]}$ for the clusters
is equal to zero within $2\sigma$, while this dependence
for the field giants differs from zero by more than
$3\sigma$ (correlation coefficients $r = -0.20 \pm 0.10$ and 
$r = -0.29 \pm 0.11$, with the probabilities of this correlation
arising by chance being $P_{N} \approx 20$\,\% and $\ll 1$\,\%,
respectively). Although the slopes for the two types of
objects coincide within the errors, the mean relative
abundances for yttrium for the clusters are higher
than those for the field stars by 
$\Delta\langle{\rm [Y/Fe]}\rangle = 0.13 \pm 0.03$

Figure~1b presents analogous diagrams for barium.
The scatter in the relative abundances of
barium is much greater for the clusters than for
the field giants, but the correlation for the clusters
is significant ($r = -0.36 \pm 0.22$, with $P_{N}<1$\,\%). 
Although the regression slopes for the clusters
are a factor of 1.5 lower than those for the
field stars ($d{\rm[Ba/Fe]}/d{\rm[Fe/H]} = -0.66 \pm 0.24$ and
$-0.98 \pm 0.09$, respectively), these two slopes are
not distinguishable due to the large uncertainties.
At the same time, the mean excess of the relative
barium abundances for clusters compared to field
giants is very high, 
$\Delta\langle{\rm [Ba/Fe]}\rangle = 0.30 \pm 0.04$. It
was concluded in [9], where a similar enhancemnet of
[Ba/Fe] was found for eight distant and mainly young
open clusters, that this enhancement is real, and is
not due to uncertainties in the abundances. The same
conclusion was drawn in [13].

This excess is equally large for lanthanum,
$\Delta\langle{\rm[La/Fe]}\rangle = 0.28 \pm 0.03$. 
The [La/Fe] values for
clusters and field giants display significant correlations
with metallicity, with the same slopes (see
Fig.~1c). Note that the scatter between the relative
abundances of this element determined in different
studies is fairly high, although the increase in the
[La/Fe] values for clusters compared to field stars
exceeds this uncertainty by more than a factor of
two; i.\,e., it is statistically significant. The relative
cerium abundances in Fig.~1d for clusters display a
very small scatter, and form a narrow sequence in
which a break near ${\rm[Fe/H]} \approx -0.2$ can even be seen.
This behavior is due to the small uncertainty in the
abundances of this element [5, Table 1]. The slopes
of the dependences for the clusters and field stars are
the same within the uncertainties, with the excess
corresponding to $\Delta\langle{\rm[Ce/Fe]}\rangle = 0.21 \pm 0.03$.

Thus, the mean relative abundances for all $s$-elements 
for any metallicity proved to be significantly
higher in clusters than in field giants. The absence of
significant systematic differences in the abundances
of the studied elements for dwarfs and giants is
demonstrated by the light gray circles in the figure
panels, representing data for field dwarfs. (The
regression fit for the dwarfs is not shown to avoid
cluttering the figure.) In spite of the fact that the
regions occupied by the field dwarfs and giants do not
always fully coincide, we can see here an enhancement
of the relative abundances of all the studied $s$-elements
for clusters compared to those for field dwarfs, for a
specified metallicity.

{\bf $r$-Elements}. According to our current understanding,
most $r$-elements are synthesized directly
during Type II supernova with masses $8 < M/M _{\odot}< 10$
(see, e.\,g., [16]). Some quantity of iron-peak elements
is also produced in Type II supernova outbursts.
Most iron-group elements are produced in
Type Ia supsernovae, which represent the final stage
in the evolution of close binary stars with masses of
$<8M_{\odot}$ (see, e.\,g., [18]). Due to the difference in the
evolution times of the stars giving rise to these two
types of supernovae, the relative abundances [$r$/Fe] in
subsequent generations of stars should decrease with
increasing metallicity. Since we are assuming that
the total heavy-element abundance in the Galactic
disk has monotonically increased over the past five
billion years [19, 20], i.\,e., that metallicity is a statistical
indicator of age, field stars of this subsystem
should display a decrease in the relative abundances
of $r$-elements with increasing metallicity, with this
decrease being the same as that for the primary 
$\alpha$-elements, since both types of element are produced
in Type II supernovae.

The metallicity dependences of the relative abundances
of europium --- the only $r$-element studied --- 
is presented in Fig.~1e. The field dwarfs and giants
form a single sequence, with their dependences fully
coincident. The regression fit constructed for all open
clusters also coincides within the uncertainties (this
fit is not shown in the figure), but the scatter in the
[Eu/Fe] values is appreciably higher for the clusters
than for the field stars. The europium abundances
for most of the clusters were determined in only one
study, but using several stars (the Eu abundances
for only 12~clusters have been determined in multiple
studies). Clusters that deviate considerably from the
band formed by the field giants are denoted in various
fonts in Fig.~1e. Clusters in which the Eu abundance
was determined in several studies or using more than
five stars in a single study are denoted in bold font,
those in which this abundance was determined using
from two to four stars in a single study in regular font,
and those in which it was determined using only one
star in a single study in italic. Clusters whose Eu
abundances were determined using only one star lie at
the periphery of the cluster of points in this diagram.
However, the position of the regression fit obtained
after excluding these clusters remains virtually unchanged.

{\bf Different groups of clusters}. Let us now compare
the elemental abundances for clusters displaying
different kinematics. We showed in [3, 4] that the
population of open clusters includes two groups with
different kinematics, as well as different chemical and
physical properties. We separated the two groups
based on the orbital parameter proposed in [21],
$(Z^{2}_{max}+4e^{2})^{1/2}$, with the division being 0.40 (for
more refined information about this boundary value,
see [5]). (Here, e is the eccentricity of the Galactic
orbit and $Z_{max}$ the maximum distance of points on
the orbit from the Galactic plane in kiloparsec.) We
referred to clusters with low orbital parameters as
``Galactic'' and to those with high orbital parameters
as ``peculiar''. We also placed clusters with thin disk
metallicities that were uncharacteristically low for
field stars (${\rm [Fe/H]} \approx< -0.2$) in the peculiar group.
However, there are no clusters with low, circular
orbits that are simultaneously metal-poor in our
sample of 90 clusters, since all such clusters are
distant and too young to contain giants that could
be used for spectroscopic determinations of their
elemental abundances. Here, all the clusters with
low orbital parameters are metal-rich (and they are
Galactic), while those with high orbital parameters
are peculiar.

The peculiar clusters are shown by hollow circles
and the Galactic clusters by circled points in Fig.~1.
The behavior of the $s$-element relative abundances
in the first four panels of this figure do not display
any significant differences between these groups of
clusters. Both groups have equally large scatters
in the [$s$/Fe] values and, on average, equally large
enhancements of these values over those for the field
stars. The $r$-element relative abundances for the
different groups of clusters, shown in Fig.~1e, behave
quite differently. The Galactic clusters lie mainly in
the band occupied by field stars (the Eu abundance
for NGC~2632 was determined for five stars in a
single study, and the Eu abundance for Mel~111 was
determine using only one star in a single study).

It is striking that the [Eu/Fe] values for all the
peculiar clusters lie below the regression fit for the
field stars (which is the mean line for these stars)
in the region ${\rm [Fe/H]} > -0.1$. The Eu abundances
for all these clusters were determined using several
stars, sometimes from several studies, making them
fairly trustworthy. The metal-poor clusters, which
mainly belong to the peculiar group, occupy a range of
[Eu/Fe] that is symmetrically broader than the band
for the field stars. Some clusters have very low and
others very high [Eu/Fe] values, not characteristic for
field stars. Note, however, that the parameters for
the three clusters Be~31, NGC~2266, and NGC~2158,
which display the strongest deviations from the mean
[Eu/Fe], were determined from only one star in a
single study. The Eu abundances of the remaining
clusters lying far from the mean values were determined
using several stars. As a result, the linear
regression fit for clusters with high, elongated orbits
has an appreciably more negative slope than the slope
for the field stars (($r = -0.70 \pm 0.03$, with $P_{N} \ll 1$\,\%).
This slope remains unchanged if we exclude clusters
with less trustworthy [Eu/Fe] values.

We showed in [5] that metal-poor clusters
(${\rm [Fe/H]} < -0.1$) with high, elongated orbits have
lower relative abundances of primary $\alpha$-elements
(oxygen and magnesium) than do field stars of the
Galactic thin disk. One possible explanation is
that the star-formation rate in the interstellar matter
far from the Galactic plane from which the high-velocity
clouds stimulating the birth of these clusters
is lower than the typical rate in the thin disk. In
this case, the relative abundances of Eu should also
be reduced, since $r$-elements are also ejected in
Type II supernovae. This difference could also come
about if the masses of the Type II supernovae that
enriched the medium from which the clusters formed
were lower, since the output of primary $\alpha$-elements
decreases with decreasing mass of the pre-supernova.
However, the relative abundances of Eu in such
clusters should then be enhanced relative to those
in field stars of the same metallicity, since $r$-elements
are produced in supernova outbursts of less massive
stars ($8 - 10M_{\odot}$), than those producing $\alpha$-elements
($>10M_{\odot}$). The large scatter in the [Eu/Fe] values in
Fig.~1e for metal-poor clusters with high, elongated
orbits could testify to the joint action of both of these
effects.

The [O,Mg/Eu] ratio is a more sensitive indicator
of the masses of Type II supernovae. We therefore
show in Fig.~1f a plot of metallicity versus the abundance
of primary $\alpha$-elements relative to Eu. The
regions occupied by the field giants and dwarfs also
coincide in this diagram. Clusters that lie far from
the region occupied by field stars are indicated. Only
three of these have their Eu abundances determined
from only one star, while the rest have Eu abundances
determined from several stars. The abundances of
primary $\alpha$-elements for all the clusters except for
Be~31 were likewise determined using several stars. It
is clear that the linear regression fits for clusters with
high, elongated orbits and for field stars differ outside
the uncertainties. Excluding the most discrepant
points in the diagram does not reduce the slope for
the peculiar clusters.

The [O, Mg/Eu] ratios for the metal-poor clusters
lie on either side of the mean line for the field stars, and
both of the mechanisms noted above could be responsible
for their large scatter, as well as possible systematic
errors in the abundances determined in various
studies. In addition, a contribution could be made by
inhomogeneity of the medium from which the high-velocity
clouds formed, and uncertainty in the fraction
of matter they contribute during the formation of the
clusters. In the positive direction, the [O, Mg/Eu]
ratios of the clusters do not extend beyond the band
occupied by the field stars, suggesting these clusters
formed from medium with an admixture of matter
ejected from supernovae with the same masses as
those exploding in the thin disk. However, the [O,
Mg/Eu] ratios deviate much more in the negative
direction in Fig.~1f, with the abundances of all elements
for the most discrepant clusters (NGC~2243,
Tombaugh~2, and Melotte~71) being determined from
several stars, making them fairly reliable. This means
that the most likely origin for these deviations is
a lower mass for the Type II supernovae that enriched
the matter from which these clusters formed,
compared to their masses in the plane of the disk.
At the same time, all the metal-rich clusters with
high, elongated orbits display not only low values of
[Eu/Fe], but high values of [O, Mg/Eu] as well. This
likely testifies that such clusters formed from mixed
material that was enriched by Type II supernovae with
very high masses.

\section {AGE DEPENDENCES OF THE RELATIVE
ABUNDANCES OF NEUTRON-CAPTURE
ELEMENTS}. 

{\bf $s$-Elements}. The age dependences of the relative
$s$-element abundances for the clusters and field
giants behave very differently (Fig.~2a-2d). Recall
that we estimated the ages of the red giants from
their masses, determined from models for their atmospheres.
The fact that the age-metallicity relations
for our red giants [5, Fig.~3a] and field dwarfs [20]
essentially coincide confirms that the ages of the red
giants are not subject to appreciable systematic errors
associated with uncertainty in the modeling. The
ages of the dwarfs were obtained using two independent
methods: from theoretical isochrones and from
their chromospheric activity.

We can see that the metallicity in the Galactic disk
was essentially independent of age during the first
several billion years, while it began to monotonically
increase in the last four to five billion years. Because
it was possible to calculate only a lower limit for the
ages of giants whose masses were given in the initial
sources with accuracy to only one significant figure
after the decimal point, the diagram constructed using
these values is discrete, and contains nine age
intervals. The corresponding step is several billions
of years for high ages, i.\,e., for low masses, and decreases
to tens of millions of years for the youngest
giants. It was easiest to compare the behavior of
the age dependences of [el/Fe] for the open clusters
and field giants by approximating the latter using line
segments joining the mean ratios $\langle{\rm [el/Fe]}\rangle$ in these
nine age intervals, and approximating the behavior for
the clusters using a smoothed trend obtained using a
sliding average.

All the plots in Fig.~2 clearly show a substantial
excess of $\langle{\rm [el/Fe]}\rangle$ for the clusters compared to the
field giants (the smallest excess is that for yttrium).
Although none of the correlations for the field giants
are formally signficant, all the ratios [el/Fe] (apart
from [Ba/Fe]) overall seem to increase slightly with
age. On the contrary, the relative abundances for
yttrium and cerium for the clusters show significant
anti-correlations (in both cases, $P_{N} < 1$\,\%). Weak
tendencies for a decrease with age for the clusters
are also displayed by barium and lanthanum, but
the correlations for these elements are formally not
significant (in both cases, $P_{N} > 15$\,\%). (The fitted regression
lines are not included in the graphs, to avoid
cluttering the figures.) At low ages, all the $s$-element
abundances for the clusters display [$s$/Fe] values significantly
higher than the solar values, while all these
ratios are close to the solar values at higher ages (the
ratios for yttrium and cerium are even slightly lower
than solar). The mean relative abundances of all the
$s$-elements for the field giants apart from barium are
significantly lower than solar at low ages, while they
are solar within the uncertainties at higher ages. The
mean values  $\langle{\rm [Ba/Fe]}\rangle$ for the 
field giants are equal to
zero within the uncertainties at all ages (the relative
abundances of barium are higher than those of the
remaining $s$-elements for both the field giants and the
open clusters).

Comparing the relative abundances of yttrium and
lanthanum in stars in young clusters and in field stars
of the thin disk, Mishenina et\,al. [9, 10] concluded
that the age dependences of [Y/Fe] and [La/Fe] for
these two types of objects were similar. They took observed
differences between the young objects within
$\delta{\rm [Y/Fe]}$, $\delta{\rm [La/Fe]} \approx 0.1$ 
to be insignificant, 
and suggested the appreciable excesses in [La/Fe] observed
for some clusters were due to differences between the
values obtained in different studies, or the possible
influence of an age variation of the metallicity resulting
from the iron output from Type Ia supernovae.
They also note that a final resolution of this question
requires additional study.

Our yttrium abundances for 34~clusters younger
than one billion years averaged over many studies
yield $\langle{\rm [Y/Fe]}\rangle = +0.07 \pm 0.02$ , 
while 75 field giants of the same age display 
$\langle{\rm [Y/Fe]}\rangle = -0.06 \pm 0.01$. Lanthanum
did not show any age dependence (Fig.~2c);
however, overall, the clusters with La abundances
have $\langle{\rm [La/Fe]}\rangle = +0.12 \pm 0.02$, 
while the field red giants have 
$\langle{\rm [La/Fe]}\rangle = - 0.13 \pm 0.01$. In other words,
there is an obvious difference in these mean values
that lies far outside the uncertainties, making
them statistically significant and impossible to ignore.
Mishenina et\,al. [9] noted the relative abundances
of barium in clusters to be enhanced relative to field
stars, especially at low ages, in full consistency with
our results.

{\bf $r$-elements}. The age dependences of the relative
abundance of the $r$-element Eu for the clusters and
field giants behave somewhat differently than the ratio
[$\alpha$/Fe], although both elements are ejected in Type II
supernovae. Figure~2e shows that the mean [Eu/Fe]
ratios for the two types of object do not depend on
age for ages $< 2$~billion years, and have 
$\langle{\rm [Eu/Fe]}\rangle \approx 0.07$, slightly 
higher than the solar values. At higher
ages, the mean ratios for the field giants begin to
grow to $\approx 4.5$ billion years, after which they remain
constant. (Although this growth is modest, the correlation
coefficient for the [Eu/Fe]--age dependence
differs from zero beyond the uncertainties.) The mean
ratios $\langle{\rm [Eu/Fe]}\rangle$ for the clusters are not significant,
but systematically decrease, becoming lower than
those for the field giants at ages $> 2$ billion years.

Figure~2f presents age dependences for the [O,
Mg/Eu] ratios for these same objects. The ratios for
the field stars smoothly increase until about three billion
years, then cease to vary with age. The mean line
for all the clusters behaves roughly similarly. All the
Galactic clusters (apart from two located very low in
the diagram) are located in the band occupied by the
field stars. Supposing that the primary $\alpha$-elements
and $r$-elements are produced in Type II supernovae
with different masses, the increase in [O, Mg/Eu]
with age for the field giants suggests that the masses
of Type II supernovae were higher in the initial stages
of the formation of the thin disk than at the current
epoch.

The metal-poor clusters with high, elongated orbits
can be divided into two groups. One has [O,
Mg/Eu] ratios that increase with age slightly more
steeply than the field stars, with their sequence lying
below the sequence for the field giants. The other,
which contains five clusters with ages in the range
0.5--3 billion years, has higher [O, Mg/Eu] ratios
than field stars of the same age. In other words, it
appears that the metal-poor group of clusters with
high, elongated orbits is made up of two subgroups.
The metal-rich clusters with high, elongated orbits
include representatives of all ages; half are located in
the band occupied by the field stars, while the others
are located in the uppermost region in the figure.

\section {RELATIONSHIP BETWEEN THE RELATIVE ABUNDANCES
OF NEUTRON-CAPTURE ELEMENTS AND THE SPATIAL POSITIONS
AND ELEMENTS OF THE GALACTIC ORBITS}.

The abundances of essentially all elements in field
stars of the Galactic thin disk decrease with increasing
Galactocentric distance. (The barium abundance
is an exception, as it does not display any radial gradient
in Cepheids [22].) In [5], we considered the radial
and vertical gradients of the metallicity and [$\alpha$/Fe]
ratios for clusters and field stars, which showed differences
between these two types of object. Figure 3a
presents dependences of [$r$/Fe] ([Eu/Fe]) on Galactocentric
position for the open clusters, Cepheids, and
field giants. (We used the apogalactic radii of their
orbits $R_a$ for the field giants, since they are all currently
located near the Sun.) This diagram shows that
the relative abundances of $r$-elements in field stars
and clusters increase with increasing Galactocentric
distance (the correlation coefficients for both types of
field stars differ from zero beyond the uncertainties,
while the correlation for the clusters is not significant
due to the large scatter in the [Eu/Fe] values, 
$P_{N} \approx 27$\,\%).

A negative metallicity gradient arises for field
stars of the thin disk because of the decrease in the
star-forming rate with distance from the Galactic
center [19]. The iron abundance in the interstellar
medium increases with time when the star-forming
rate decreases in a stellar-gas system, mainly due
to ejection from comparatively long-lived progenitors
of Type Ia supernovae. Therefore, like [$\alpha$/Fe], [$r$/Fe]
should increase with increasing $R_G$, both currently
and in subsequent generations of stars, as is observed
for Eu in Fig.~3a (and for [$\alpha$/Fe] in Fig.~4b of [5]).
On the other hand, open clusters (especially those
with high, elongated orbits) are genetically related
to each other only weakly, and radial gradients can
arise in them only if they acquired at the protocloud
phase an admixture of matter with a different chemical
composition, which simultaneously conveyed an
impulse that elongated their initially circular orbits.
The existence of a correlation between the metallicity
and distance from the Galactic center for the clusters
testifies that the amount of admixed metal-poor
matter they acquired from high-velocity clouds is
proportional to the associated impulse acquired [5].

We found in [5] that $\alpha$-elements in clusters display
a slightly lower slope in the dependence of [$\alpha$/Fe] on
$R_G$ (see Fig.~4b of [5]) than field giants and Cepheids.
At the same time, the radial gradients of the relative
abundances of Eu for the clusters, field Cepheids,
and field giants in Fig.~3a are very similar. However,
corresponding regression fits for the peculiar clusters
in Fig.~3a display a significantly higher slope than
the slope for both types of field stars. Thus, the
Galactocentric-distance dependences for the relative
abundances of Eu for the peculiar clusters with high,
elongated orbits and the field stars are substantially
different.

In order to correct for possible systematic errors in
the abundances for each element, we further consider
the behavior of averaged abundances for two $s$-elements
(Ba and La). (Unfortunately, the abundances
of the different $s$-elements were determined for stars in
different clusters, and we therefore used mean ratios
for these two $s$-elements for all the objects, since they
were most often determined for the same clusters.)
The relative abundances of $s$-elements in Fig.~3b show
that the radial gradients obtained for the open clusters
based on their current positions are much greater,
and the sequence of values itself much higher, than
for the field giants, based on the apogalactic radii
of their orbits. The slope of the [$s$/Fe]--$R_a$ relation
does not change if we adopt the apogalactic radii of
the cluster orbits instead of their current positions.
(On the other hand, for reasons we do not currently
understand, the [$s$/Fe]--$R_G$ relation and associated
gradients for the clusters and field Cepheids proved
to be very similar.) All the corresponding correlations
are significant ($P_{N} < 5$\,\%).

The output of $s$-elements depends on the metallicity
of the stars that eject them, and is not related to
the history of star formation in the stellar-gas system.
Therefore, the presence of the observed slopes of the
[$s$/Fe] dependences on the Galactocentric positions
for the field giants in Fig.~3b can be explained by
the existence of a negative radial metallicity gradient
in the Galactic disk, while we suggest that the
cluster dependence is due to the relationship between
the amount of metal-poor matter they acquired and
the velocity impulse imparted to them by the high-velocity
clouds that stimulated their birth.

Figures~3c,~3d present the dependences of the
relative abundances of the $r$ and $s$ elements in clusters
on the orbital parameter $(Z^{2}_{max}+4e^{2})^{1/2}$. (Five clusters
with implausibly high orbital parameters, $> 3.0$,
which were probably a result of errors in their spatial
velocities, were excluded.) The lines in all the panels
show the regression lines for peculiar clusters with
$(Z^{2}_{max}+4e^{2})^{1/2} > 0.40$. These plots show that the
dependences of the abundances of all elements considered
on this kinematic parameter are formed by the
clusters with high, elongated orbits. As our analysis
showed, the points in the orbit that are maximally
distant from the Galactic plane $Z_{max}$ make a large
contribution to the orbital parameter, and not the
orbital eccentricity. This figure shows that the slopes
of the dependences for the field giants and clusters
coincide within the uncertainties in both diagrams;
while the relative abundances of Eu for the clusters
are, on average, lower for any orbital parameters, the
[$s$/Fe] ratios are much higher.

\section {DISCUSSION} 

According to [4], only Galactic clusters -- metal-rich
clusters (${\rm [Fe/H]} >\approx -0.2$) with flat, circular orbits
($(Z^{2}_{max}+4e^{2})^{1/2} < 0.40$) --- could form from the
same material as field stars. Therefore, it would
seem that the relative abundances of their neutron-capture
elements should lie in the band occupied by
field stars in plots of the metallicity dependences of
these abundances. We showed in [5] that the relative
abundances of $\alpha$-elements satisfy this hypothesis.
This is also more or less true for the relative abundances
of the $r$-element Eu (Fig.~1e). However, the
relative abundances of $s$-elements in Galactic clusters
demonstrate substantial excesses relative to field
stars.

The dependences for all the neutron-capture elements
for clusters with high, strongly elongated orbits
differ from those for field stars. Since $r$-elements
and primary $\alpha$-elements are ejected into the interstellar
medium in Type II supernovae, this suggests
that they should display the same metallicity dependences.
In reality, the slopes of plots of [Eu/Fe] versus
[Fe/H] for peculiar clusters are significantly higher
than those for field stars (Fig.~1e), while the fitted
regression line for the relative abundances of primary
$\alpha$-elements for clusters [5, Fig.~2c] has an appreciably
lower slope than the dependence for the field giants.
Thus, metal-rich and metal-poor clusters with high,
elongated orbits display differences from field giants
with the same metallicity in both [O, Mg/Fe] and
[Eu/Fe]. The [O, Mg/Eu] ratios for these two groups
of clusters with high, elongated orbits also proved
to be very different. All the metal-rich clusters have
higher values of these ratios than field giants, on average.
The metal-poor clusters display enhanced and
reduced ratios of primary $\alpha$-elements to Eu (although
remaining within the region occupied by field giants);
however, among the latter, several clusters deviate far
behond the sequence of the field stars.

The age dependences of the relative abundances of
the $r$-element Eu in clusters and field stars also appear
very different: [Eu/Fe] (like [$\alpha$/Fe]) increases with
age for the field giants, while these ratios monotonically
decrease with age for the clusters (Fig.~2e). Although
[$\alpha$/Fe] increases with age for the clusters, this
increase is appreciably slower than the increase for
field stars. As a result, the mean relative abundances
of both $\alpha$-elements and Eu for clusters and field stars
coincide at low ages; at high ages, the [$\alpha$/Fe] and
[Eu/Fe] ratios for clusters become appreciably lower.

At all Galactocentric distances, a substantial
fraction of clusters with high, elongated orbits have
[Eu/Fe] values lying below the band occupied by
the field stars (Fig~3c). The [Eu/Fe] dependence on
the orbital parameter also lies far lower for clusters
with high, elongated orbits than for field giants, and
the two regression slopes differ significantly. As a
result, the relative abundances of Eu are different
for clusters with high, elongated orbits and field
stars. The differences between metal-rich and metal-poor
peculiar clusters are also clear. In particular,
the more metal-poor group has higher 
$\langle{\rm [Eu/Fe]}\rangle$ values, on average, 
corresponding to an appreciable
slope in a plot of [Fe/H] versus [Eu/Fe]. This same
difference in the mean relative abundances of Eu led
to the existence of a substantial radial gradient for
the peculiar clusters, and a dependence on the orbital
parameter.

The properties of the relative abundances of $s$-elements
in open clusters and field stars of the thin disk
also differ substantially. Although the [$s$/Fe] ratios for
both types of object decrease with increasing [Fe/H]
in approximately the same way, the cluster values are
substantially higher than those for the field stars for
any metallicity (Fig.~1). The [$s$/Fe] ratios for the field
Cepheids are similarly high. We proposed in [14] that
the high relative abundances of $s$-elements in young
Cepheids could have come about because collapsing
high-mass supernovae ceased to explode in the thin
disk several hundreds of millions of years ago, as a
result of an increase in the overall metallicity. As a
result, the relative abundances of elements ejected by
Type II supernova were redistributed in the interstellar
medium. However,this explanation is not suitable
for clusters, since many of the clusters for which
elemental abundances are available have high ages,
sometimes comparable to the age of the Galactic
thin disk, and relative abundances of $s$-elements that
are high for a given metallicity are also displayed by
clusters with fairly low [Fe/H] values.

In contrast to $\alpha$-elements and $r$-elements, $s$-elements
display high relative abundances compared
to field stars at ages $\leq 2$~billion years, while their mean
values become similar at high ages (Fig.~2). [$s$/Fe]
increases with increasing Galactocentric distance for
clusters, with the dependences for the clusters and
field Cepheids fully coinciding. The radial gradient
in the relative abundances of $s$-elements is slightly
smaller for field giants, and the sequence itself lies
much lower (Fig~3b).

The [$s$/Fe] ratio for clusters also increases with
the orbital parameter, while this ratio is essentially
independent of the shape of the orbit for the field stars.
As a result, these elements likewise testify to the
absence of a genetic relationship between field stars
of the thin disk and open clusters. It was concluded
in [9, 10], where the same method was used to obtain
the relative barium abundances in several young open
clusters and in field stars, that the origin of the higher
[Ba/Fe] values for clusters was not associated with
systematic errors. It is not yet clear how to explain
this result. It appears that the same situation occurs
with the relative abundances of the remaining $s$-elements
in open clusters.

\section {CONCLUSION}

Thus, the characteristic differences in the behavior
of the relative abundances of primary $\alpha$-elements and
$r$-elements in open clusters support the conclusion,
based on an analysis of their metallicities and the
elements of their Galactic orbits, that metal-poor
clusters (${\rm [Fe/H]} < - 0.1$) with high, elongated orbits
were formed during interactions of metal-poor high-velocity
clouds and the interstellar material of the
Galactic thin disk. The lower relative abundances of
primary $\alpha$-elements, and frequently lower [O, Mg/Eu]
ratios, in these clusters compared to field stars, can
most logically be understood if the masses of the
Type II supernovae that enriched the material from
which these high-velocity clouds formed more often
fell in the range $(8 - 10)M_{\odot}$, than is the case for the
supernova that enriched the material of the thin disk.
The large scatter in the relative abundances of Eu
in metal-poor clusters with high, elongated orbits
suggest that this material was poorly mixed.

On the other hand, the higher [O, Mg/Fe] and
[O, Mg/Eu] ratios in clusters with high, elongated
orbits and ${\rm [Fe/H]} > -0.1$, compared to field stars,
suggests that these clusters were born as a result
of interactions of matter of the thin disk with high-velocity
clouds formed in a ``Galactic fountain'' process.
Such clouds are believed to form from matter
ejected from the thin disk as a result of massive supernova
explosions, after which this material falls back
toward the disk, stimulating the birth of open clusters
in some cases. The detection of higher relative
abundances of primary $\alpha$-elements and lower relative
abundances of $r$-elements in metal-rich clusters with
high, elongated orbits, on average, compared to field
stars, apparently testifies that such Galactic fountains
form mainly as a result of the explosion of more
massive Type II supernovae ($M > 10M_{\odot}$), for which
the output of primary $\alpha$-elements is higher, and the
output of $r$-elements lower, than is the case for less
massive supernovae.

A distinguishing property of the relative abundances
of $s$-elements in open clusters (independent
of the group to which they belong) is their appreciable
excesses compared to field stars of the thin
disk. Only for very old clusters ($t > 4$~billion years)
with high, elongated orbits the [$s$/Fe] ratios are the
same as those for field stars, within the uncertainties,
while they are substantially higher in younger clusters,
independent of the nature of the clusters. Since
most clusters are young, on average, clusters display
enhanced relative abundances of all $s$-elements,
compared to field stars. This is not inconsistent with
the proposed origin of clusters with high, elongated
orbits presented above, but it is not possible to place
them in a logical, self-consistent scheme. There is no
question that there remains only possible differences
of their properties from those for the same elements
in field stars of the Galactic thin disk. The possible
origins of this behavior for the $s$-elements is not yet
understood.

\section*{ACKNOWLEDGMENTS}

The authors thank N.O. Budanova for help in collecting
the material on the chemical compositions of
open cluster stars.
This work was supported by the Ministry of Education
and Science of the Russian Federation (state
contracts 3.961.2014/K, 213.01-11/2014-5 (project
code 26.63)), and the Southern Federal University
(grant 213.01-2014/013-VG).

\renewcommand{\refname}{Список литературы}

\newpage

\begin{figure*}
\centering
\includegraphics[angle=0,width=0.87\textwidth,clip]{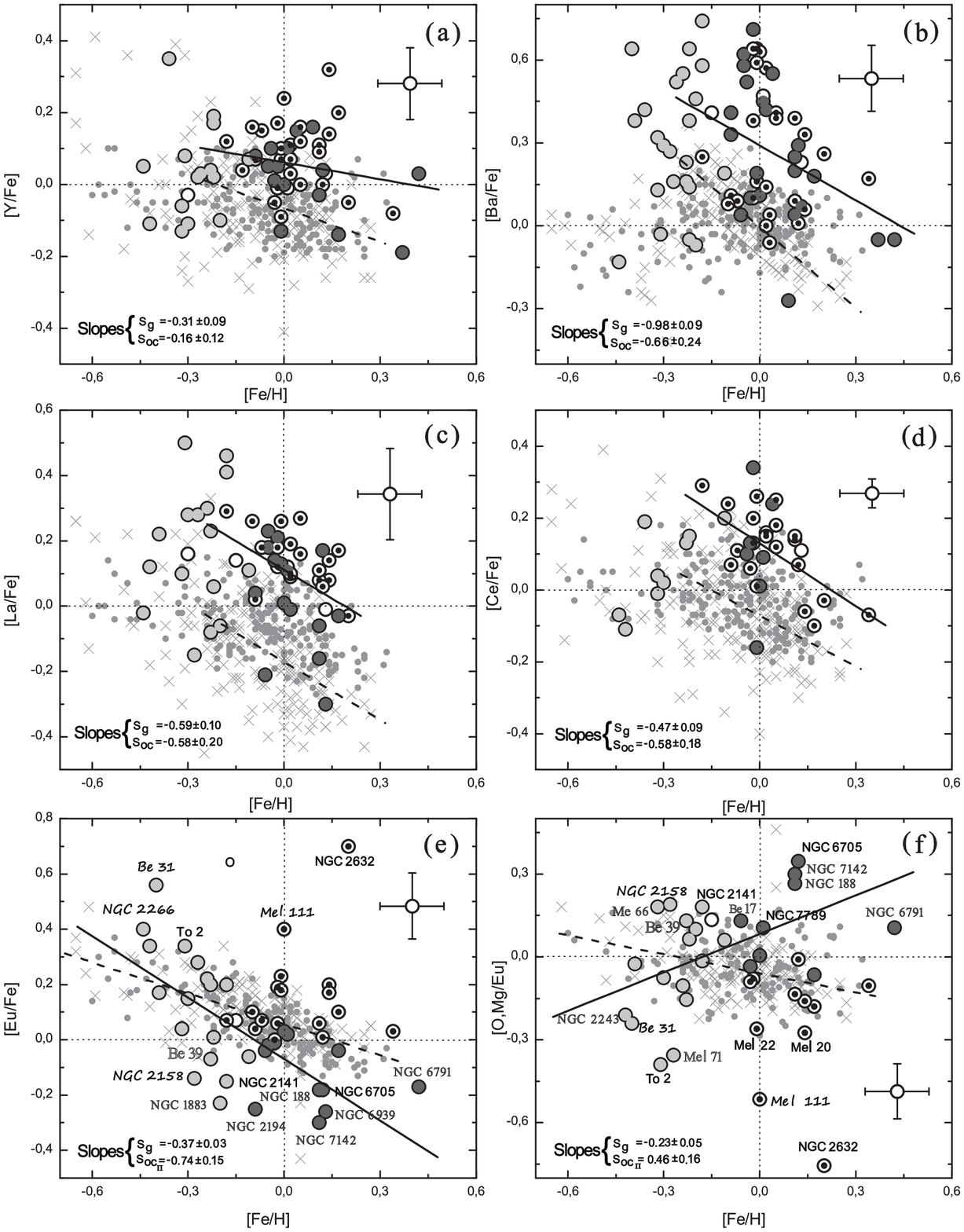}
\caption{Metallicity dependences of the relative abundances of (a)--(d) 
         four $s$-elements, (e) an $r$-element, and (f) [O, Mg/Eu] in
         open clusters (circles), field red giants (x's), and field 
         dwarfs (gray points). The circled dots show clusters with low, 
         circular orbits (orbital parameter 
         ($(Z^{2}_{max}+4e^{2})^{1/2} < 0.40$ and $|z| < 400$~pc). 
         The gray circles show clusters in high, elongated orbits,
         with light gray circles corresponding to metal-poor 
         (${\rm [Fe/H]} < -0.1$) and dark gray circles to metal-rich 
         clusters. The hollow circles show unclassified clusters. 
         The solid and dashed lines in panels (a)--(d) show the results 
         of regression fits for the clusters and giants, respectively, 
         and the solid and dashed lines in panels (e)--(f) for clusters 
         with high, elongated orbits and field giants, respectively. 
         The slopes for the regression fits are shown in the panels 
         with their errors. The bars show the characteristic
         uncertainties for all the elements. The dotted lines parallel 
         to the coordinate axes are drawn through the solar values. 
         The names of clusters that deviate strongly from the region 
         occupied by field giants are indicated in panels (e) and (f); 
         those in bold font have oxygen,magnesium, and europium 
         abundance determinations in several studies or for more than 
         five stars in a single study, those in regular font have them 
         for two to four stars in a single study, and those in italic 
         have them for one star in a single
study.}
\label{fig1}
\end{figure*}

\newpage

\begin{figure*}
\centering
\includegraphics[angle=0,width=0.99\textwidth,clip]{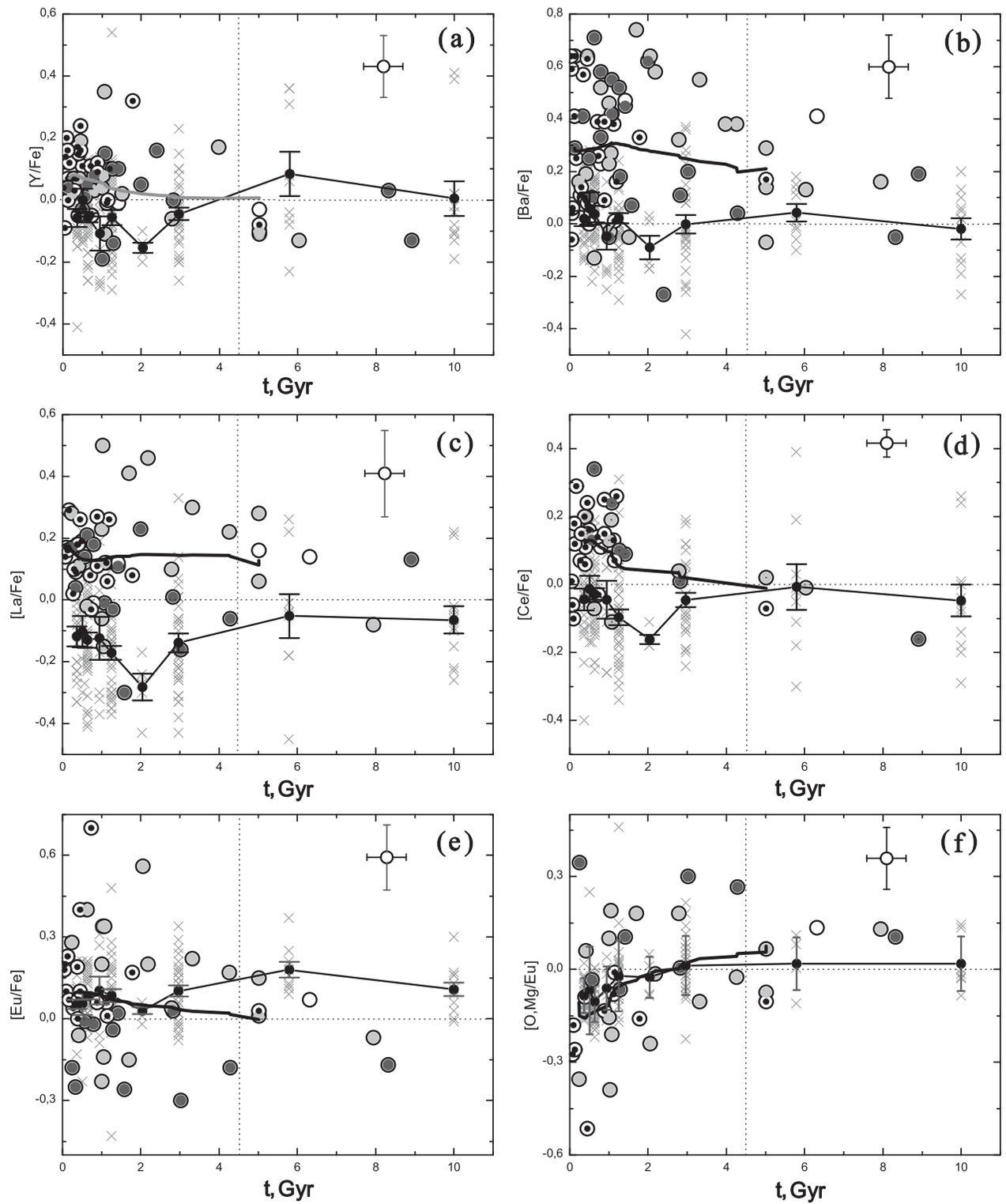}
\caption{Age dependences of the relative abundances of (a)--(d) 
         four $s$-elements, (e) one $r$-element, and (f) [O, Mg/Eu] 
         for open clusters (circles) and field red giants (x's). 
         The smooth curves show fits to the smoothed trends obtained 
         using a sliding average of the age dependences of the 
         clusters; the curves with jointed line segments show the mean 
         metallicities and $\langle [s/{\rm Fe}]\rangle$ values in nine 
         age ranges for the field giants. The bars indicate the mean 
         errors for the giants. The remaining notation is as in
Fig. 1.}
\label{fig2}
\end{figure*}

\newpage

\begin{figure*}
\centering
\includegraphics[angle=0,width=0.95\textwidth,clip]{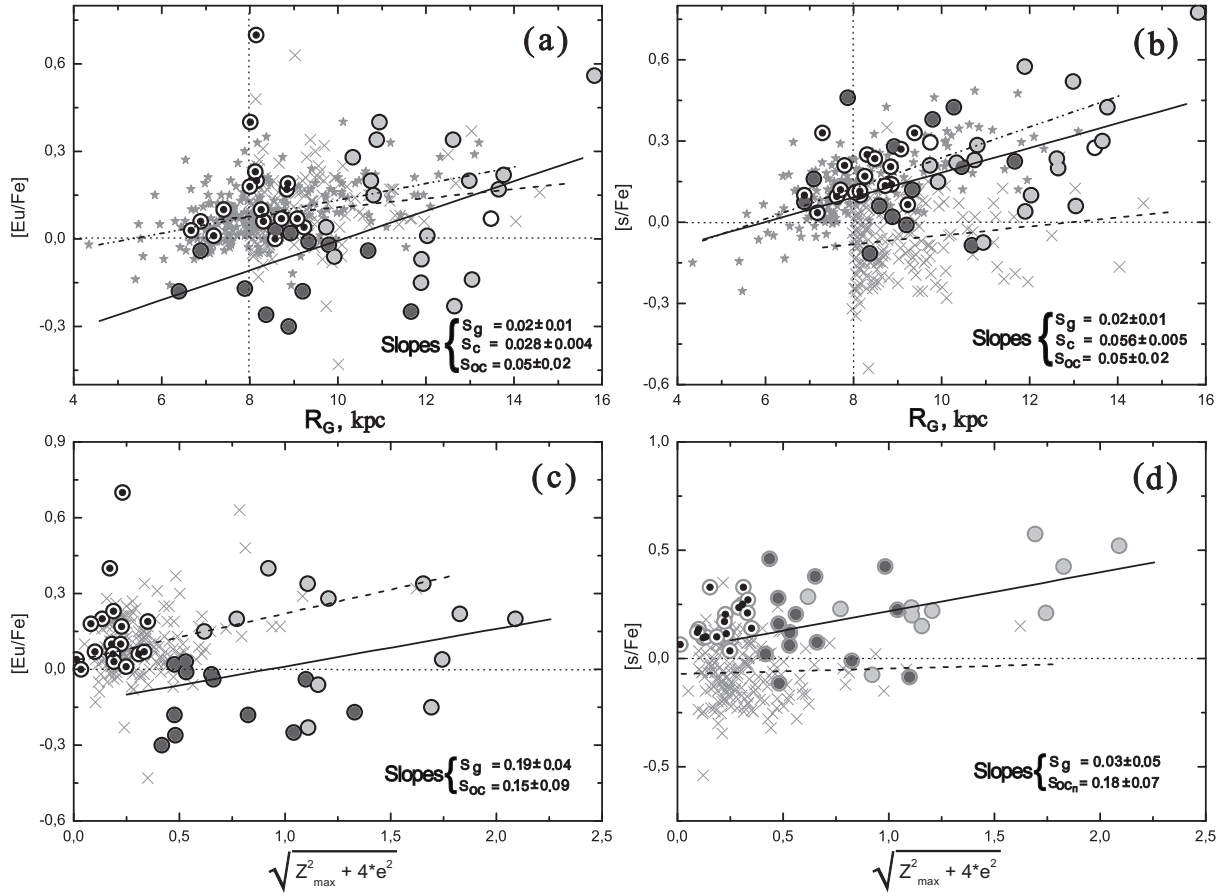}
\caption{Dependences of the relative abundances of an $r$-element 
         (Eu) and two averaged $s$-elements (La, Ba) on (a)--(b)
         Galactocentric position and (c)--(d) the orbital parameter 
         $(Z^{2}_{max}+4e^{2})^{1/2}$ for open clusters (circles), 
         field red giants (x's), and field Cepheids (stars). In view 
         of the nearness of the field giants to the Sun, we used the 
         apogalactic radii of their orbits in these plots. The solid, 
         dashed, and dot-dashed lines show the regression fits for the 
         peculiar clusters, field giants, and field Cepheids, 
         respectively. The remaining notation is as in Fig.~1.}
\label{fig3}
\end{figure*}

\end{document}